# Luttinger Liquid Behavior in a Single-Layer Nickelate $La_{1.4}Sr_{0.6}NiO_4$

Yinqi Hu[1], Yidian Li[1], Xian Du[2], Wenxuan Zhao[1], Kaiyi Zhai[1], Senyao Zhang[1], Jiawei Shao[1], Mingxin Mao[1], Houke Chen[3], Jieyi Liu[3,4], Donghui Lu[5], Makoto Hashimoto[5], Fangyuan Zhu[6], Zhengtai Liu[6], Dawei Shen[7], Yaobo Huang[6], Zhongkai Liu[8,9], D. Prabhakaran[3], Yanpeng Qi[8,9,10†], Yilin Wang[11,12,13†], Yulin Chen[3,8,9†], Lexian Yang[1,14†]

*[1]State Key Laboratory of Low Dimensional Quantum Physics, Department of Physics, Tsinghua University, Beijing 100084, China*

*[2]Department of Applied Physics, Yale University, New Haven, CT 06511, USA*

*[3]Department of Physics, Clarendon Laboratory, University of Oxford, Parks Road, Oxford OX1 3PU, UK*

*[4]Diamond Light Source, Harwell Science and Innovation Campus, Didcot OX11 0DE, UK*

*[5]Stanford Synchrotron Radiation Lightsource, SLAC National Accelerator Laboratory, Menlo Park, CA 94025, USA*

*[6]Shanghai Synchrotron Radiation Facility, Shanghai Advanced Research Institute, Chinese Academy of Sciences, Shanghai 201210, China*

*[7]National Synchrotron Radiation Laboratory and School of Nuclear Science and Technology, University of Science and Technology of China, Hefei 230026, China*

*[8]School of Physical Science and Technology, ShanghaiTech University, Shanghai 201210, China*

*[9]ShanghaiTech Laboratory for Topological Physics, Shanghai 200031, China*

*[10]Shanghai Key Laboratory of High-resolution Electron Microscopy, ShanghaiTech University, Shanghai 201210, China*

*[11]School of Emerging Technology, University of Science and Technology of China, Hefei 230026, China*

*[12]New Cornerstone Science Laboratory, University of Science and Technology of China, Hefei, 230026, China*

*[13]Hefei National Laboratory, University of Science and Technology of China, Hefei 230088, China*

*[14]Frontier Science Center for Quantum Information, Beijing 100084, China*

*Emails: LXY: lxyang@tsinghua.edu.cn; YLC: yulin.chen@physics.ox.ac.uk; YPQ: qiyp@shanghaitech.edu.cn; YLW: yilinwang@ustc.edu.cn*

**Abstract:**

**The discovery of high-temperature superconductivity in bilayer and trilayer nickelates has spurred intense interest in the Ruddlesden-Popper nickelates; yet the fundamental properties of the $NiO_2$ layer remain obscured by interlayer coupling. It is therefore imperative to investigate the electronic properties of their single-layer counterpart — though not superconducting — to isolate the intrinsic physics of the NiO layer. In this work, we present a systematic study of the single-layer nickelate $La_{1.4}Sr_{0.6}NiO_4$ using high-resolution angle-resolved photoemission spectroscopy (ARPES) and theoretical calculations. We reveal strong electron correlation effects, manifested by high-energy kinks in band dispersions and a pronounced orbital-dependent band renormalization. Interestingly, we observe a quasi-one-dimensional electronic structure characterized by straight Fermi surface sheets along the diagonal momentum directions. Such square Fermi surface topology facilitates non-Fermi liquid behavior consistent with the Luttinger liquid model, as evidenced by the power-law spectral function, robust temperature scaling, and the observation of spin-charge separation. Our results therefore not only unveil an exotic Luttinger liquid behavior emerging from the unexpected dimensional reduction in an intrinsically quasi-two-dimensional nickelate but also provide a new perspective for understanding the intriguing physics in multilayer nickelates.**

## I. INTRODUCTION

Nickelates, long regarded as structural and electronic analogues of cuprates, have recently emerged as a compelling platform for exploring high-temperature superconductivity [1-7]. In particular, layered nickelates in Ruddlesden-Popper (RP) phase demonstrate unconventional superconductivity, with critical temperature exceeding the liquid nitrogen boiling point [8-15]. These breakthroughs have spurred a surge of investigations, aiming at unravelling the unique electronic mechanisms that underpin their superconducting and strongly correlated behaviors [16-18]. Among the many theoretical and experimental findings, the non-Fermi liquid behavior of the RP-phase nickelates, such as $La_3Ni_2O_7$ and $La_4Ni_3O_{10}$, is of particular interest [19-26]. Although the origin of this anomalous behavior remains elusive, its resolution is enssential to advancing our understanding of the electronic landscape and the nature of unconventional superconductivity in RP-phase nickelates.

Notably, Fermi surfaces characterized by large flat portions have been widely observed in various nickelate systems, including bilayer $La_3Ni_2O_7$ and $La_4Ni_3O_{10}$ [23,27,28], the alternating monolayer-trilayer $La_3Ni_2O_7$ [29], and $LaNiO_3$ thin films [30,31]. Such Fermiology is reminiscent of quasi-one-dimensional (1D) electronic systems [32,33], except that the parallel flat segments in these compounds run along two orthogonal momentum directions, forming a square-like geometry. Theoretical studies have explored the implications of the square Fermi surface based on a square lattice model dominated by nearest-neighbor hopping (Fig. 1a and 1b), suggesting its potential connection to density-wave instability and unconventional superconductivity [34-36]. The square geometry is particularly intriguing, as it gives rise to non-Fermi liquid behavior [34-36], resonating with Anderson’s hypothesis that the 2D electron gas behaves like the Luttinger liquid (LL) in 1D rather than the Fermi liquid [37]. Nevertheless, experimental validation of this connection remains sparse. Nickelates offer a rare opportunity to explore this intriguing problem. However, the strong interlayer interaction and octahedron distortion in multi-layer nickelates complicate the Fermi surface structure and obscure the intrinsic physics of the Ni-O square lattice. Consequently, their single-layer counterpart offers an optimal route to disentangle the pivotal role of Ni-O planes and interlayer interaction in shaping the electronic ground states, motivating a focused interrogation of the intrinsic physics within the Ni-O layer.

In this work, we present a systematic study of the single-layer RP-phase nickelate $La_{1.4}Sr_{0.6}NiO_4$ using angle-resolved photoemission spectroscopy (ARPES) and *ab initio* calculations to elucidate the intrinsic physics of the Ni-O layer. We observe strong electron correlation effects, manifested by high-energy kinks resembling the band dispersion of cuprates, alongside pronounced orbital-dependent band renormalization resulting from strong Coulomb interaction and Hund's coupling. Prominently, our work reveals a quasi-1D electronic structure along orthogonal momentum directions, forming flat Fermi surface segments. Such a square Fermi surface engenders the non-Fermi liquid behavior that follows the LL model, as evidenced by a power-law spectral function, robust temperature scaling behavior, and the observation of spin-charge separation [34-36]. Our findings therefore demonstrate an exotic LL behavior, which originates from an unexpected reduction in the dimensionality of the electronic structure within a quasi-two-dimensional system. These results establish single-layer nickelate as a compelling platform for investigating the interplay among electron correlation, Hund's coupling, and dimensionality in multi-orbital electronic system [38-40], which further provide important insights for deciphering anomalous electronic properties of multi-layer nickelates.

## II. RESULTS

Figure 1c schematically illustrates the crystal structure of $La_{2-x}Sr_xNiO_4$ in the RP phase. It comprises alternating layers of corner-sharing $NiO_6$ octahedra and La/Sr atomic layers, periodically stacked along the *c* axis. The out-of-plane tilt of the $NiO_6$ octahedra is suppressed by replacing La with Sr. With *x* larger than 0.1, $La_{2-x}Sr_xNiO_4$ crystallizes in the space group I4/mmm [41]. As presented in Fig. 1d, the square Ni-O plane is similar to the Cu-O plane in cuprates, with nearest-neighbor Ni atoms connected by in-plane O atoms. Figure 1e shows the 3D Brillouin zone and its surface projection with high-symmetry points indicated. The Fermi surface of $La_{1.4}Sr_{0.6}NiO_4$ calculated by density-functional theory (DFT) consists of nearly parallel sheets extending along $\bar{\Gamma}\bar{X}$ (Fig. 1f), in contrast to that of $La_3Ni_2O_7$ with rounded-square or circular Fermi pockets in the DFT calculation [23,42]. The energy bands crossing $E_F$ are contributed by $d_{x^2-y^2}$ and $d_{z^2}$ orbitals (Fig. 1g), similar to $La_3Ni_2O_7$ and $La_4Ni_3O_{10}$ [23,27]. Nevertheless, the $d_{z^2}$ orbital does not show a flat band but quickly disperses across $E_F$. Figure 1h displays temperature-dependent resistivity under selected pressures. At low temperatures, $La_{1.4}Sr_{0.6}NiO_4$ exhibits insulating behavior without any indication of superconductivity up to 69 GPa, which underscores the critical role of interlayer hopping in the emergence of superconductivity in bilayer and trilayer nickelates [42-46]. At 2.67 GPa, the resistivity curve shows a kink around 220 K, which may be related to charge or spin ordering [47,48].

Figure 2 presents the basic band structure of $La_{1.4}Sr_{0.6}NiO_4$ measured with high-resolution ARPES. The constant-energy contours at different binding energies consist of parallel line segments along $\bar{\Gamma}\bar{X}$ (Fig. 2a). While in overall agreement with the DFT calculation in Fig. 1f, the experimental Fermi surface seems to be straighter, resembling the electronic structure of quasi-1D electronic systems [32,33]. The Fermi surface of $La_{1.4}Sr_{0.6}NiO_4$ can be regarded as two sets of parallel lines with weak hybridization at the crossing points (Fig. 2b), which is different from those of $La_3Ni_2O_7$ and $La_4Ni_3O_{10}$ with a circular electron pocket around $\bar{\Gamma}$ (also see the Supplemental Materials [49]). The invariance of the momentum distribution curves (MDCs) along the Fermi surface (indicated by the white rectangle in Fig. 2b) confirms the straightness of the parallel Fermi surface sheets (Fig. 2c). Moreover, the Fermi momenta and Fermi velocity show minor dependence on the momentum (Fig. 2d), reinforcing the quasi-1D character of the electronic structure of $La_{1.4}Sr_{0.6}NiO_4$.

Figures 2e-h present the band dispersions along high-symmetry directions. We observe two bands crossing $E_F$, which are contributed by the $d_{x^2-y^2}$ and $d_{z^2}$ orbitals, according to the DFT calculations (Fig. 2e). In contrast to $La_3Ni_2O_7$ and $La_4Ni_3O_{10}$, the top of the $d_{z^2}$ band is far above $E_F$, leaving no flat band near $E_F$. In the experiments, the α band derived from $d_{x^2-y^2}$ orbital disperses down to -1 eV, below which the spectra are dominated by strong and slowly dispersing features. The DFT-calculated band structures are overlaid for comparison, showing overall agreement between the experiments and calculated results. The α band is well reproduced by the calculation, while a substantial renormalization by a factor of about 3 is required for the calculated β band that is dominated by the $d_{z^2}$ orbital character near the $\bar{\Gamma}$ point, suggesting an orbital-dependent electron correlation effect (see the Supplemental

Materials [49]), in line with multi-layer nickelates $La_3Ni_2O_7$ and $La_4Ni_3O_{10}$. Compared to $La_{1.4}Sr_{0.6}NiO_4$, the band renormalization in $La_3Ni_2O_7$ and $La_4Ni_3O_{10}$ is enhanced roughly by a factor of 2 due to interlayer interaction. Given the comparable Ni valence, this points to an important role of interlayer electron correlation effect in the physics of multi-layer compounds [50].

Interestingly, the band dispersion along $\bar{M}\bar{X}$ (Fig. 2g), comprising a mixture of the two orbitals (see Supplemental Materials VII [49]), shows a pronounced kink-like structure around -260 meV, as better visualized in the zoom-in plots in Fig. 3a and 3b. Such a "waterfall"-like feature in the band dispersion has been commonly observed in many correlated materials and is arguably regarded as a manifestation of strong electron correlation [51-59]. By fitting the momentum-distribution curves (MDCs) at different binding energies to Lorentzian functions (see the Supplemental Materials [49]), we extract the electron self-energy of $La_{1.4}Sr_{0.6}NiO_4$. The imaginary part of the electron self-energy gradually increases with the binding energy, while the real part exhibits a maximum at approximately 260 meV below the Fermi energy (Fig. 3c, d), which is far beyond the energy range of normal lattice vibrations. The electron self-energies of cuprate $Bi_{1.74}Pb_{0.38}Sr_{1.88}CuO_{6+\delta}$ (Bi2201) are replotted from Ref. [51] for comparison. The comparable onset energy of the waterfall-like feature in single-layer nickelate and cuprate suggests their similar electron correlation effects [54], although the electron correlation effect is further modulated by Hund's coupling in nickelates (see the Supplemental Materials [49]).

Having established basic electronic structure and electron correlation effects, we now focus on the spectral features in the vicinity of $E_F$. We notice that the spectral weight is significantly suppressed upon approaching $E_F$ (Fig. 2 and Fig. 3), a hallmark of non-Fermi liquid behavior. In Fig. 4, we scrutinize this non-Fermi liquid behavior based on the LL model, motivated by the quasi-1D electronic structure and square Fermi surface of $La_{1.4}Sr_{0.6}NiO_4$. Fig. 4(a) presents the temperature-dependent ARPES spectra measured using laser-ARPES with improved resolutions. Consistent with the spectra measured with synchrotron-based ARPES, no clear quasiparticle peak is observed near $E_F$, similar to the results in $La_3Ni_2O_7$ and $La_4Ni_3O_{10}$ [23,27,28]. Fig. 4b compares ARPES spectra of $La_{1.4}Sr_{0.6}NiO_4$ with those of prototypical Fermi liquid and LL materials. $La_{1.4}Sr_{0.6}NiO_4$ is clearly distinct from the polycrystalline gold, a reference Fermi liquid with a well-defined Fermi edge. Instead, it closely resembles the prototypical quasi-1D LL materials [32,33,60].

To quantify the LL behavior of $La_{1.4}Sr_{0.6}NiO_4$, we fit the experimental data to the LL model at finite temperature [33]:

$$\rho(\epsilon, T) \propto T^{\alpha} Re\left[(2\mathrm{i})^{\alpha+1} B\left(\frac{\alpha+1-\frac{\mathrm{i}\epsilon}{\pi}}{2}, -\alpha\right)\right],$$

Where $\rho\,(\epsilon,\,T)$ represents the LL spectral function, $\epsilon = (E - E_F)/k_B T$ is the scaled energy with $k_B$ the Boltzmann constant. $B$ is the beta function, and α represents the LL anomalous exponent. A system is considered a Fermi liquid when $\alpha = 0$, whereas it becomes a LL for $\alpha > 0$. As shown in Fig. 4c, the ARPES data of $La_{1.4}Sr_{0.6}NiO_4$ fit perfectly to the LL model over a wide temperature range from 21 K to 185 K, yielding a temperature-independent exponent $\alpha \approx 0.39 \pm 0.04$ (see the Supplemental Materials [49]). The LL property can be independently verified through its characteristic scaling behavior with temperature. That is, the integrated ARPES spectra scaled by $T^{\alpha}$ should be temperature independent, which is excellently verified in Fig. 4d with $\alpha \approx 0.37$. Furthermore, the LL model predicts a linear dependence of $\log\,[I(E_F)]$ on $\log T$ with the slope corresponding to the LL exponent $\alpha$, where $I$ represents the spectral intensity. This linear relationship with $\alpha \approx 0.38$ is demonstrated in Fig. 4e, consistent with both the model fitting and the scaling analysis. Remarkably, we observe two band dispersions at the $\bar{X}$ point, compared to one band predicted by DFT calculation, as better visualized by the curvature plot (Fig. 4f). This result is in close analogy to the previous observation of spin-charge separation [61-64]. We therefore assign these two bands to separated dispersion branches of spinon and chargon. Based on the above experimental findings, we compellingly identify the LL properties of $La_{1.4}Sr_{0.6}NiO_4$, which is associated with the quasi-one-dimensional electronic structure hidden in its nominally two-dimensional crystal structure.

## III. DISCUSSION

The observation of LL behavior in a strongly correlated square lattice is beyond expectation, as the LL model is typically restricted to quasi-1D systems. While LL physics has been reported in other quasi-two-dimensional materials, those systems usually feature quasi-1D structural motifs [33,65]. In η-$Mo_4O_{11}$, for example, LL physics is preserved by orthogonal orbital components along different chain directions. This scenario, however, is not applicable to $La_{1.4}Sr_{0.6}NiO_4$, where the LL behavior emerges within an intrinsically two-dimensional Ni-O plane, and the electrons along orthogonal directions exhibit mixed orbital characters. Remarkably, our observation closely conforms to the theoretical model of square Fermi surfaces [34-36]. The square geometry of the Fermi surface suggests that nearest-neighbor hopping dominates the low-energy electronic structure. This is consistent with the tight-binding model on a square lattice, where the Fermi surface becomes perfectly nested in the absence of further-neighbor hopping. Our DFT calculations and ARPES measurements both indicate that the electronic bands near the Fermi level are primarily derived from the Ni $d_{x^2-y^2}$ and $d_{z^2}$ orbitals, with hopping integrals dominated by nearest-neighbor terms (see the Supplemental Materials for the details about the hopping integrals extracted through Wannier simulation of the band structure [49]). This simple model nicely captures the essential quasi-1D character of the electronic excitations, facilitating the mapping of the square lattice to effective one-dimensional chains along orthogonal directions [34-36]. Within this framework, the bosonic component of electron-electron interactions destabilizes the conventional Fermi liquid phase, giving rise to LL behavior. Moreover, theoretical studies have shown that for repulsive Hubbard interactions on a square lattice with perfect nesting, density-wave instabilities (e.g., spin or charge density waves) are favored. Our findings in $La_{1.4}Sr_{0.6}NiO_4$ are consistent with this picture: the system exhibits strong correlations and a tendency toward density-wave order, as hinted by the kink in the resistivity under pressure.

Our findings in the single-layer nickelate $La_{1.4}Sr_{0.6}NiO_4$ also provide important insights into the physics of nickelates. On the one hand, the experiment isolates the intralayer and interlayer electron correlation effect. A comparison of the electronic structures across single- and multi-layer nickelates suggests that intralayer correlation effects generate similar “waterfall” features in the $d_{x^2-y^2}$ band dispersions, whereas interlayer correlation effects further renormalize the bandwidth, Fermi velocity, and carrier effective mass in multi-layer nickelates [54]. Consistent with the previous theoretical proposal [50], this result suggests a dominant role of interlayer electron correlation effect in the physics of multi-layer nickelates. On the other hand, bilayer and trilayer nickelates exhibit rounded Fermi surfaces due to interlayer hopping, which suppresses the LL behavior. This warping disrupts the perfect nesting condition inherent in the square Fermi surface, which tends to suppress density-wave instabilities and may favor superconducting pairing [34-36]. This picture resonates with P. W. Anderson's proposal that the unconventional superconductivity is possibly associated with the LL-like physics in higher dimensions. These results also provide a possible understanding of the non-Fermi liquid behavior observed in RP-phase nickelates: although the LL behavior within the NiO layer is destroyed by the interlayer coupling, it may still affect the electronic behavior in multi-layer nickelates.

In conclusion, our comprehensive investigation of $La_{1.4}Sr_{0.6}NiO_4$ reveals key information on its electronic structure and correlation effects. Our ARPES experiments provide compelling evidence for the LL behavior of $La_{1.4}Sr_{0.6}NiO_4$. By isolating the physics of the single Ni-O layer without interlayer interaction, our work not only provides important insights into the novel electronic behaviors in nickelates but also accentuates the delicate interplay between dimensionality and electronic instabilities in correlated materials.

## IV. METHOD

### A. ARPES experiment

Single crystals of $La_{1.4}Sr_{0.6}NiO_4$ were grown by the high-pressure optical floating zone technique. Synchrotron-based ARPES measurements were conducted at beamline 5-2 of Stanford Synchrotron Radiation Lightsource (SSRL) under proposal No. S-XV-ST-6370. The samples were cleaved *in-situ* under ultra-high vacuum below $5\times10^{-11}$ mbar. Data were collected with a Scienta DA30 electron analyzer. The total energy and angular resolutions were set to 15 meV and 0.2°, respectively.

Preliminary data were collected at beamline Bloch (Proposal No. 20230668) in MAX IV, beamline 03U (Proposal No. 2025-SSRF-PT-513724) and beamline 09U (Proposal No. 2025-SSRF-PT-512611) in Shanghai Synchrotron Radiation Facility (SSRF), and beamline 10.0.1 (Proposal No. ALS-11649) in Advanced Light Source (ALS).

Laser-based ARPES measurements were conducted at Tsinghua University. The 7-eV laser was delivered by frequency doubling in a KBBF crystal and then focused on the sample with a beam spot of about $10 \times 10\ \mu m^2$. Samples were cleaved *in-situ* under ultra-high vacuum below $6\times10^{-11}$ mbar. Data were collected with a Scienta DA30 electron analyzer. The total energy and angular resolutions were set to 3 meV and 0.2°, respectively.

**B. DFT+DMFT calculation**

The DFT calculations were performed using the WIEN2K package [66]. We simulate the Sr-doping effect by reducing the total valence electrons and compensating it by a uniform background of positive charges in DFT. We perform fully charge-self-consistent single-site DFT+DMFT calculations using the DFT+eDMFT code developed by Haule et al. [67,68] based on WIEN2K. We choose a large hybridization energy window from -10 to 10 eV. All the five Ni-3d orbitals are considered as correlated ones and a local Coulomb interaction Hamiltonian with Ising-form is used, which is parameterized by on-site Hubbard $U$ and Hund's coupling $J_H$. We choose the nominal double-counting scheme with a nominal Ni-3*d* occupancy of 7.4 electrons, corresponding to the Sr-doping of $x = 0.6$. The continuous time quantum Monte Carlo [69] is used as impurity solver, and a total of $2.56 \times 10^9$ samplings are performed. The self-energy on real frequency is obtained by the analytical continuation method of maximum entropy. All the calculations are performed at 100 K. We use the Wannier90 package [70] to construct a tight-binding model consisting of Ni $d_{x^2-y^2}$ and $d_{z^2}$ orbitals for a monolayer $La_{1.4}Sr_{0.6}NiO_4$.

## ACKNOWLEDGMENTS

This work is funded by the National Key R&D Program of China (Grant No. 2022YFA1403201 and Grant No. 2022YFA1403100), the National Natural Science Foundation of China (No. 12274251, 12004270, 52272265, and 12174365). L.X.Y. acknowledges support from the Tsinghua University Initiative Scientific Research Program and the Fund of Science and Technology on Surface Physics and Chemistry Laboratory (No. XKFZ202102). Y.L.W. acknowledges support from the Innovation Program for Quantum Science and Technology (No. 2021ZD0302803) and the New Cornerstone Science Foundation. Use of the Stanford Synchrotron Radiation Light Source, SLAC National Accelerator Laboratory, is supported by the US Department of Energy (DOE), Office of Science, Office of Basic Energy Sciences under contract no. DE-AC02-76SF00515. This research used resources of the Advanced Light Source, a US DOE Office of Science User Facility under contract no. DE-AC0205CH11231.

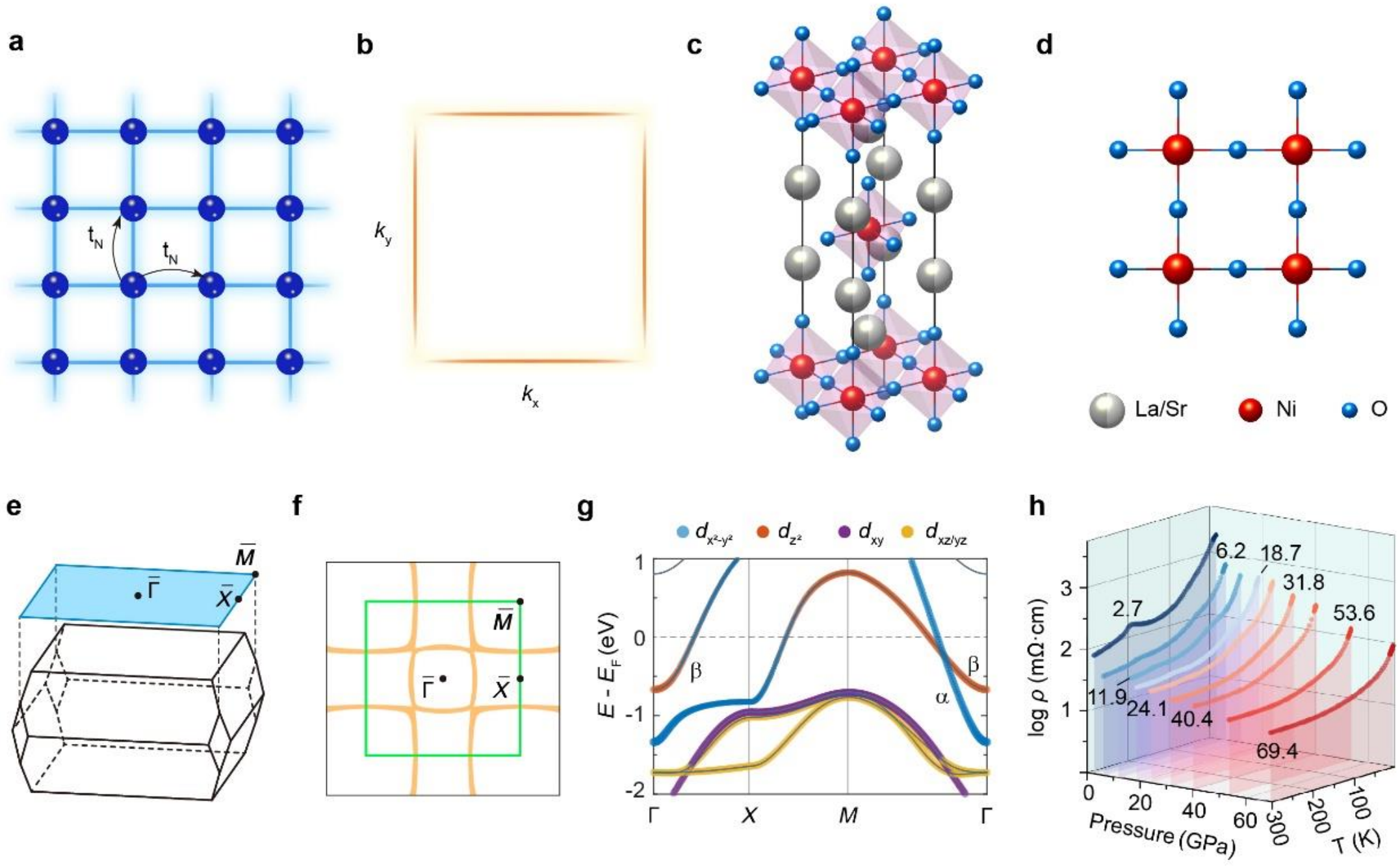

**Figure 1**. **Basic properties of $La_{1.4}Sr_{0.6}NiO_4$. a**, Schematic illustration of a square lattice with nearest-neighbor hopping dominating the low-energy electronic excitations. **b**, Schematic of the square Fermi surface of the two-dimensional (2D) lattice model in **a**. **c**, Illustration of the crystal structure of $La_{2-x}Sr_xNiO_4$ with Ni-O layers separated by La-O layers. **d**, The Ni-O layer as the basic construction unit in nickelate superconductors. **e**, 3D Brillouin zone of $La_{2-x}Sr_xNiO_4$ and its surface projection. **f**, Fermi surface (FS) of $La_{1.4}Sr_{0.6}NiO_4$ calculated by density-functional theory (DFT). **g**, DFT-calculated orbital-projected band structure of $La_{1.4}Sr_{0.6}NiO_4$. **h**, Temperature-dependent resistivity under different pressures.

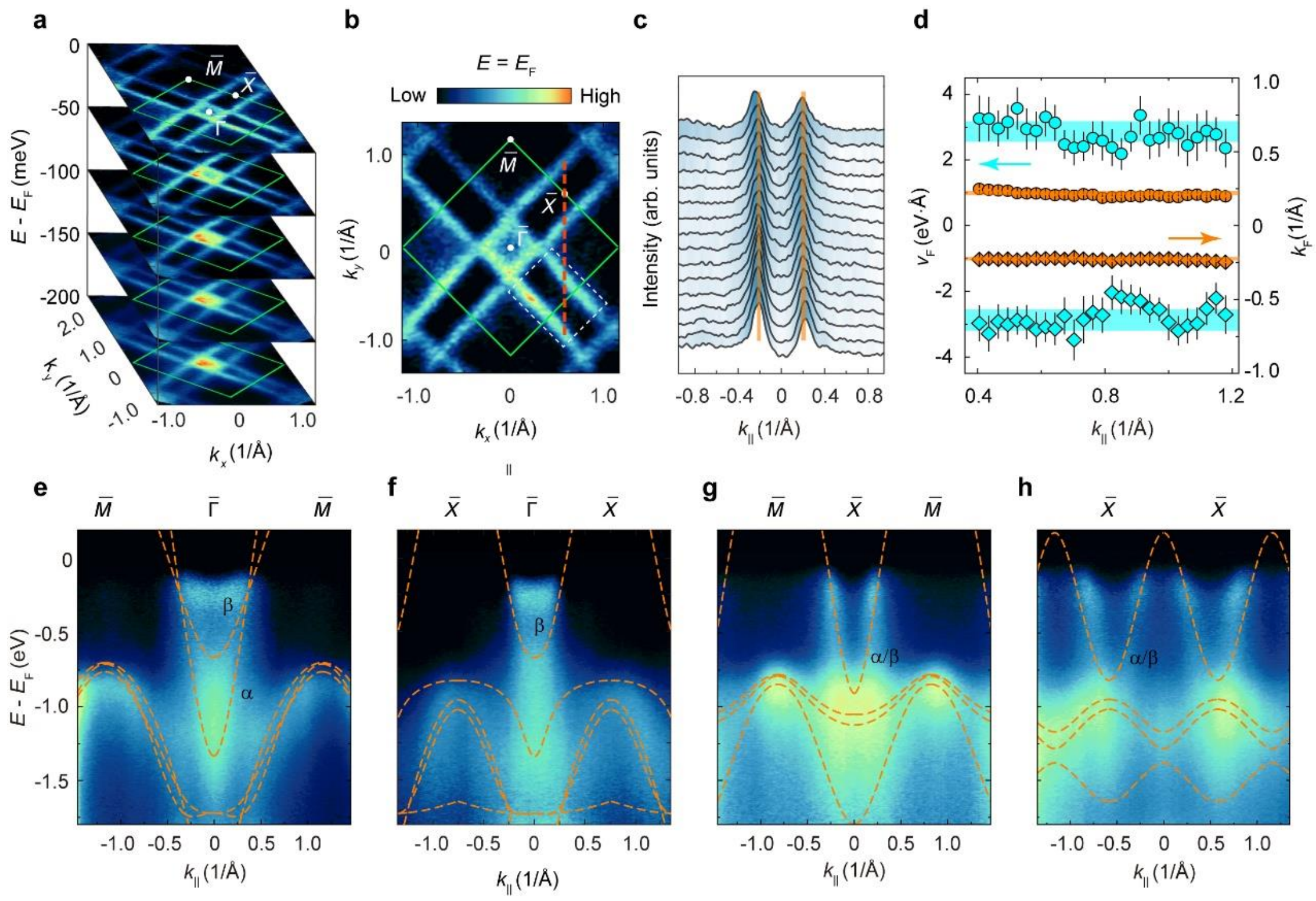

**Figure 2. Basic electronic structure of $La_{1.4}Sr_{0.6}NiO_4$. a**, Stacking plot of constant-energy contours at selected binding energies. The green lines indicate the first Brillouin zone (BZ). **b**, Fermi surface measured by integrating ARPES intensity in an energy window of 40 meV around the Fermi level ($E_F$). **c**, Stacking plot of momentum-distribution curves at $E_F$ in the momentum region as indicated by the white dashed rectangle in panel **b**. **d**, Fermi momenta and Fermi velocity as functions of momentum in the white dashed rectangle in panel **b**. **e-h**, Band dispersions along different momentum directions. The momentum direction of panel **h** is indicated by the red dashed line in **b**. DFT calculated band structures (orange dashed lines) are overlaid for comparison. Data in **e**, **h** were collected using linear-horizontally (LH) polarized photons at 153 eV, while data in **a**, **b**, **f**, **g** were collected using LH polarized photons at 182 eV. All data were collected at 15 K.

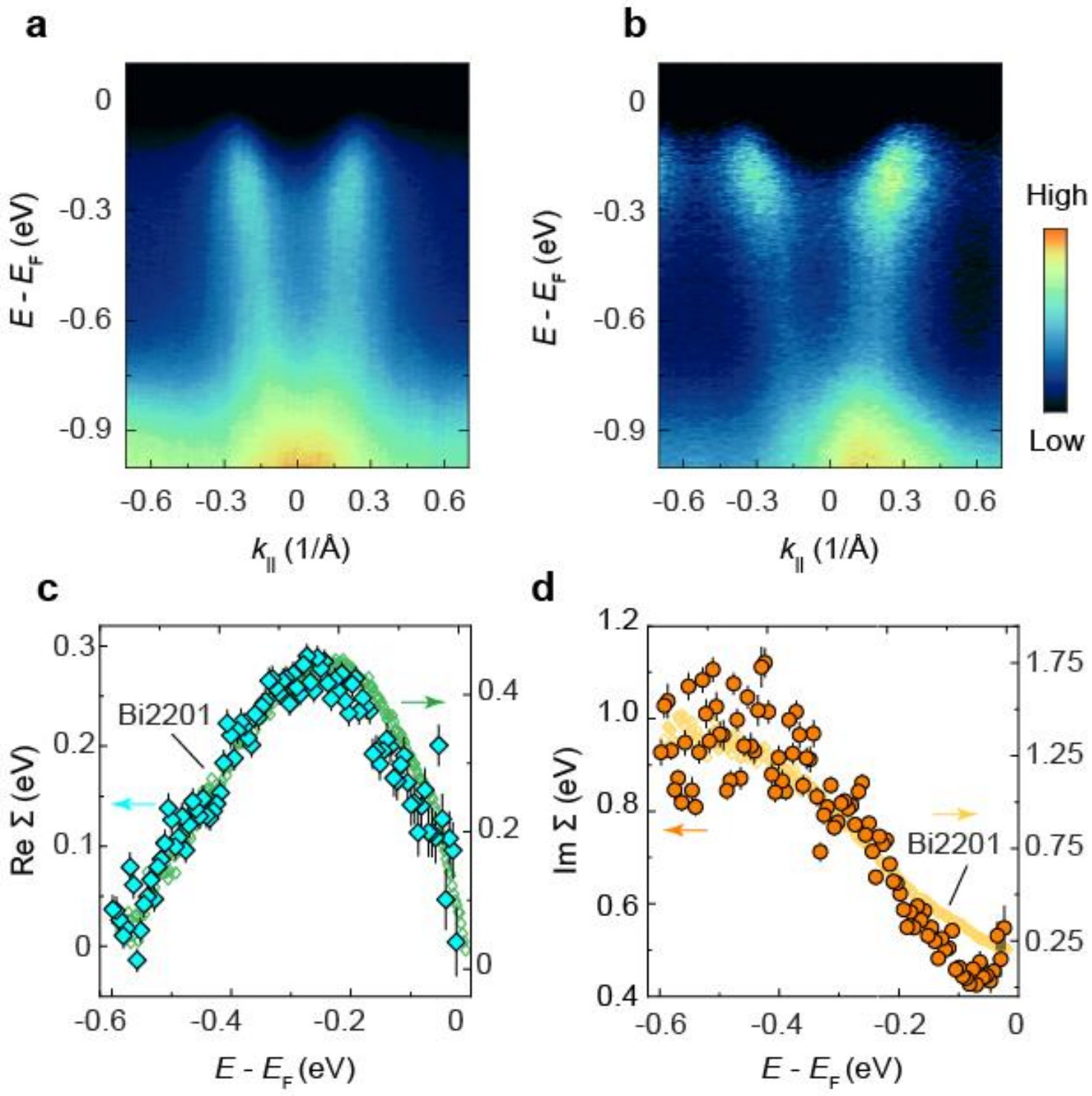

**Figure 3. Electron correlation effect of $La_{1.4}Sr_{0.6}NiO_4$. a**, **b**, Band dispersions along $\bar{X}\bar{M}$ and the red dashed line in Fig. 2b measured at 15 K. Data were collected using LH polarized photons at 163 eV and 153 eV, respectively. **c**, **d**, Real and imaginary part of electron self-energy extracted from band dispersion in (b), with real and imaginary part of Bi2201 overlaid for comparison. Data of Bi2201 were collected at 20 K with 22.5 eV photons according to Ref. [51].

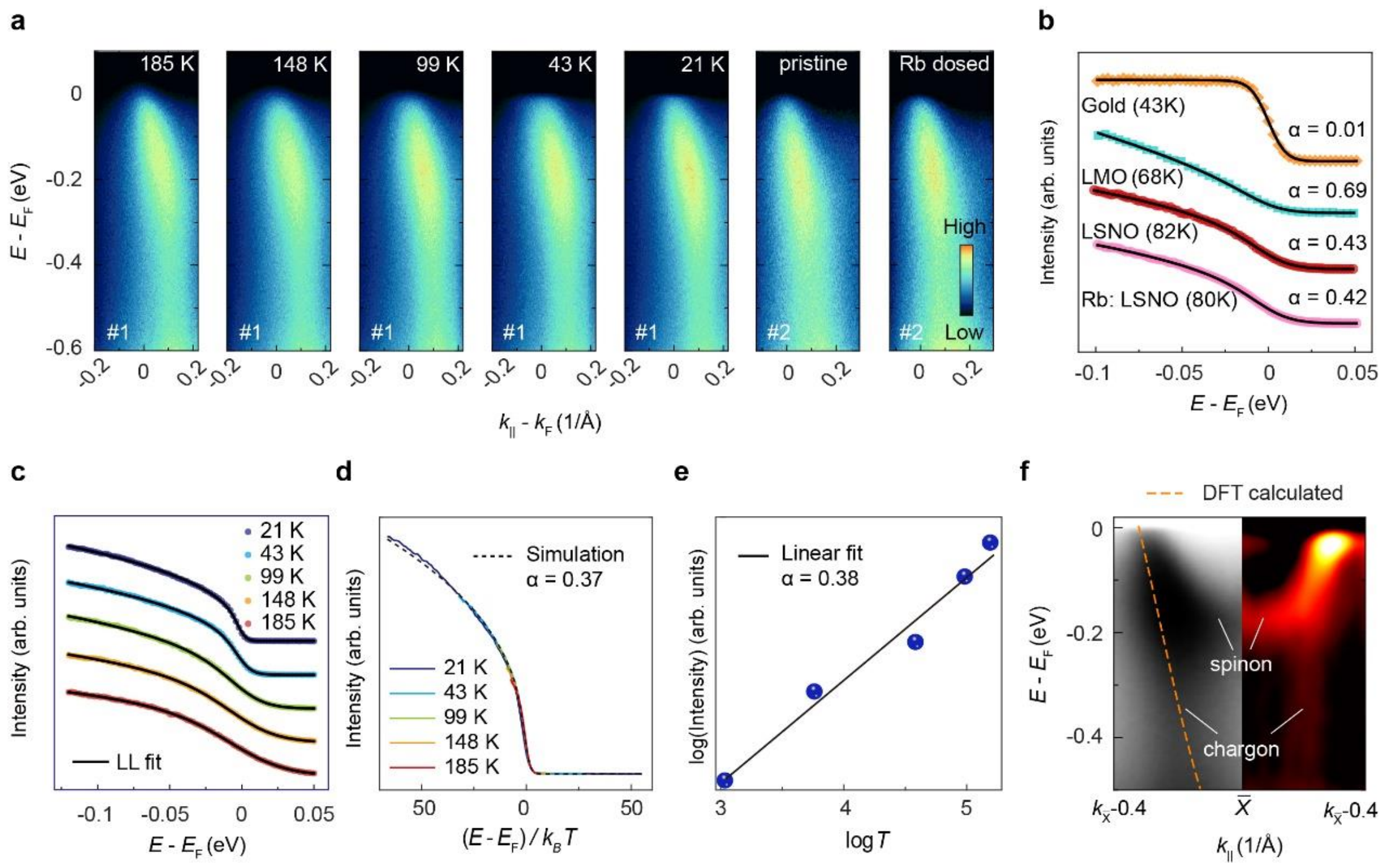

**Figure 4. Non-Fermi liquid behavior of $La_{1.4}Sr_{0.6}NiO_4$. a**, ARPES measurements of the band dispersion along $\bar{\Gamma}\bar{M}$ at selected temperatures using LH polarized laser at 7eV. Temperature-dependent data and Rb-doping dependent data were collected from sample #1 and #2, respectively. **b**, Comparison of integrated EDCs of polycrystalline gold, $Li_{0.9}Mo_6O_{11}$ and $La_{1.4}Sr_{0.6}NiO_4$ before and after Rb doping with LL exponent indicated. **c**, Integrated EDCs of $La_{1.4}Sr_{0.6}NiO_4$ measured at different temperatures and the fits to the LL model. **d**, Scaling plot of the EDCs in **c** with a scaling factor of $T^{0.37}$ (see the main text for details). The dashed line plots the simulated LL spectral function with anomalous exponent $\alpha$ = 0.37. **e**, log-log plot of the EDC intensity at $E_F$ versus temperature. The black line is a linear fit to the data with $\alpha$ = 0.38. **f**, The observation of the spin-charge separation. Left: Raw data measured at $\bar{X}$ along the red line in Fig. 2b. The orange dashed line is the DFT-calculated band dispersion. Only one

band is predicted by the calculation. Right: The curvature plot of the spectra. Two bands are clearly observed. Data were measured using 51 eV photons with LH polarization at 15 K.